\documentclass[%
reprint, amsmath,amssymb, aps,superscriptaddress]{revtex4-1}

\usepackage{graphicx}
\usepackage{dcolumn}
\usepackage{bm}
\usepackage{natbib}

\begin{document}
\preprint{APS/123-QED}
\title{A new quantity for statistical analysis: "Scaling invariable Benford distance"}

\author{Peiyan Luo}
\affiliation{ %
 College of Nuclear Technology and Automation Engineering, Chengdu University of Technology, Chengdu, China}%
\author{Yongqing Li}
 \email{liyongqing@scu.edu.cn}
\affiliation{ 
College of Physical Science and Technology, Sichuan University, Chengdu, China}%
\date{\today}
\begin{abstract}
For the first time, we introduce "Scaling invariable Benford distance" and "Benford cyclic graph", which can be used to analyze any data set. Using the quantity and the graph, we analyze some date sets with common distributions, such as normal, exponent, etc., find that different data set has a much different value of "Scaling invariable Benford distance" and different figure feature of "Benford cyclic graph". We also explore the influence of data size on "Scaling invariable Benford distance", and find that it firstly reduces with data size increasing, then approximate to a fixed value when the size is large enough.
\begin{description}
\item[PACS numbers]
02.50.Cw, 05.90.$^+$m
\end{description}
\end{abstract}
\maketitle
\section{\label{sec:level1}Introduction }
The nine digits 1--9 produced by first digit analysis of our typical
everyday numbers are supposed to be occurring randomly and thus equally
distributed. But in 1881, Newcomb~\cite{NS} 
found that the first digit 
proportions of many numbers were quite different, and in 1938, Benford~\cite{BF} 
gave the exact expression \(P_B(d)=\log_{10}(1+1/d)\), where \(P_B(d)\) is the probability of first
digit $d$ occurring in a data set. This is known as \emph{Benford'law}. Since foundation of the law, much research~\cite{AEK,MJN,AB,TP1,TP2,TP3,BB,JB,EL,MBQ1,MBQ2,MBQ3,JRI,RG}
 has been done on giving theoretical extensions, analyzing more cases, or applying the law to other fields. Apparently, not all of the data sets have the same first digit distribution
 as \emph{Benford'law}. Moreover, the data set with finite size results in rational numbers of first digit proportions, while \emph{Benford'law} gives irrational numbers, in other words, the data set will never exactly meet the logarithmic law. Meanwhile, we usually
 don't know how the first digit proportions change with the data size increasing, and we also don't know the limit of the proportions.

The question is how much the difference between the first digit proportions of a data set and the first digit law is, and it's the focal issue of this paper. Considering that a data set satisfying \emph{Benford'law} is scaling symmetric, which means that it still satisfies the law when
members of the data set multiply an arbitrary number. And if the data set is not 
consistent with the law, it is scaling asymmetric. So, we propose a new quantity which does not vary with units or scales, such quantity is so necessary because that
units and scales of data set are artificially chosen at all time. That is "Scaling invariable Benford distance" we'll talk about in the latter part. 
\section{\label{sec:level2} "Scaling invariable Benford distance"}
Let $A_{0}$ be a data set with finite size. Firstly, an distance $I[A_0]$ is defined 
using \emph{Benford'law},
\begin{equation} \label{eq:Ia0}
I[A_0]=\sqrt{\sum_{d=1}^9[P_{A_0}(d)-P_B(d)]^2}
\end{equation}
Where $P_{A_0}(d)$ is the probability of first digit $d$ occurring in $A_0$.
Then, a new data set $A_{\alpha}$ is given by transforming $A_0$,
 \(A_{\alpha}=\{x_\alpha|x_\alpha=x_0\times10^\alpha,x_0\in A_0 \}\), where $\alpha$ is a random number between 0 and 1. And $I[A_\alpha]$ can be similarly defined as $I[A_0]$, 
\begin{equation} \label{eq:Iaa}
I[A_\alpha]=\sqrt{\sum_{d=1}^9[P_{A_\alpha}(d)-P_B(d)]^2}
\end{equation}
Where $P_{A_\alpha}(d)$ is the probability of the occurrence of first number $d$ in $A_\alpha$. Generally, $I[A_0]\neq I[A_\alpha]$.

Now, we define a new quantity $I_{\textrm{inva}}[A_\alpha]$ in Eq.(~\ref{eq:Iinva}), and name it "Scaling invariable Benford distance",
\begin{equation} \label{eq:Iinva}
I_{\textrm{inva}}[A_\alpha]=\int_{0}^1 I[A_{\alpha+\beta}]d\beta
\end{equation}
Obviously, 
\(I_{\textrm{inva}}[A_\alpha]=I_{\textrm{inva}}[A_\beta]=I_{\textrm{inva}}[A_0] \), where $I_{\textrm{inva}}[A_\beta]$ and $I_{\textrm{inva}}[A_0]$ have the same definition as $I_{\textrm{inva}}[A_\alpha]$. That is to say, $I_{\textrm{inva}}[A_\alpha]$ is a fixed value which is independent of $\alpha$.

Thus, for any data set $A_{\alpha}$, there is a quantity--"Scaling invariable Benford distance" which doesn't change with units or scales. And the central question is tackled through this quantity displaying the difference between the first digit proportions of a data set and \emph{Benford'law}.
\section{\label{sec:level3}Analysis of data sets}
\subsection{\label{sec:level31} Data sets satisfying and approximating \emph{Benford'law}}
The data set $X$ showed in Eq.(~\ref{eqx}) is  consistent with \emph{Benford'law},
{\setlength\arraycolsep{2pt}
\begin{eqnarray} \label{eqx}
X & = & \{x|x=10^{\beta}, \nonumber\\
  &   & \beta \in[0,1)\textrm{ with uniform distribution} \} 
\end{eqnarray}}
And $Y$ is an arbitary data set,
\begin{equation} \label{eqy}
Y=\{y|y \, \textrm{ is nonzero and arbitrary number} \}
\end{equation}
Then, data set $Z$ can be given from $X$ and $Y$,
\begin{equation} \label{eqz}
Z=\{z|z=x\times y,x\in X,y\in Y\}
\end{equation}
For nonzero $Y$ with any distribution, data set $Z$ satisfies the first digit law, which can be simply proved latter in a different way compared to Hamming~\cite{HM}.

Firstly, $Y$ can be rewritten as
{\setlength\arraycolsep{2pt}
\begin{eqnarray} \label{eq:Y}
Y & = & \{y(\beta_0,t)|y(\beta_0,t)=10^{\beta_0+N(t)}, \nonumber\\
     &   &\beta_0 \in [0,1),N(t)\,\textrm{is an integer} \} 
\end{eqnarray}}
where only $\beta_0$ contributes to the first digit.
\\
Then,
{\setlength\arraycolsep{2pt}
\begin{eqnarray}
Z & = & \{z|z=10^{\beta+\beta_0+N(t)}\}  \nonumber \\
  & = & \{z|z=10^{\tilde{\beta}+N(t)},\tilde{\beta} \in [\beta_0,\beta_0+1)\} \nonumber \\
  & = & \{z|z=10^{\tilde{\beta}+N(t)},\tilde{\beta} \in [\beta_0,1)\}\bigcup  \nonumber \\
  &   & \{z|z=10^{\tilde{\beta}+N(t)},\tilde{\beta} \in  [1,\beta_0+1)\} \nonumber \\
  & = & \{z|z=10^{\tilde{\beta}+N(t)},\tilde{\beta} \in [\beta_0,1)\} \bigcup \nonumber \\
  &   & \{z|z=10^{\tilde{\beta}+N(t)+1},\tilde{\beta} \in [0,\beta_0)\}  
\end{eqnarray}}
In addition, data set $X$ can also be expressed as
{\setlength\arraycolsep{2pt}
\begin{eqnarray} \label{eq.X}
X & = & \{z|z=10^{\beta},\beta \in [\beta_0,1)\} \bigcup \nonumber \\
  &   &  \{z|z=10^{\beta},\beta \in [0,\beta_0)\}  
\end{eqnarray}}
Thus, data set $Z$ has the same distribution of first digit as $X$. That is to say, $Z$ satisfies \emph{Benford'law}.

\begin{figure}[b]
\includegraphics[width=0.5\textwidth]{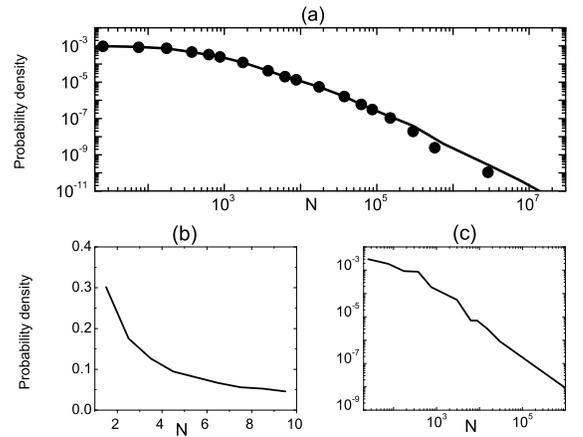}
\caption{\label{fig:fig1} Distributions of the data sets, i.e (a) "pop-2009"~\cite{AEK} (line) and "pop-c" (dot), (b) $X$ defined in Eq.~\ref{eqx}, (c) $Y$ produced by us}
\end{figure}

Notice that data sets like $Z$ are a large category of data sets for the arbitrariness of $Y$. Thus, a real-world data set, which approximates \emph{Benford'law}, may be 
a data set like $Z$, for example, the USA Census Data "pop-2009" ~\cite{AEK}.
Here, we can produce a data set named "pop-c" with the same distribution as "pop-2009", where "pop-c" is given from two data sets, one is $X$ defined in Eq.~\ref{eqx} and another is $Y$ with distribution showed in graph (c) of FIG.~\ref{fig:fig1}. And distributions of "pop-2009", "pop-c" and $X$ are also showed in FIG.~\ref{fig:fig1}.
%
\subsection{\label{sec:level32}Analyzing some common data sets}
Now, we analyze some typical data sets, distribution functions of which are showed in TABLE~\ref{tab1}, and these four cases are labeled as \emph{Normal}, \emph{Exponent}, \emph{Uniform} and \emph{Constant} respectively. Members of any case here are random numbers generated by computer to form the data set, the size of which is set to be $10^5$.
\begin{table}
\caption{\label{tab1} Distribution functions }
\begin{tabular}{l| l}
\hline
\emph{Normal}    &   $p(x)=\frac{1}{\sqrt{2\pi}}e^{-x^2/2},x\in(-\infty,\infty)$  \\
\hline
\emph{Exponent}  &   $p(x)=e^{-x},x\in[0,\infty)$  \\
\hline
\emph{Uniform}  &   $p(x)=1/9,x\in[1,10]$  \\
\hline
\emph{Constant} &   $p(x)=1,x=9$ \\
\hline
\end{tabular}
\end{table}

For each data set, we give the results of "Scaling invariable Benford distance"  $I_{\textrm{inva}}[A_\alpha]$ [Eq.(~\ref{eq:Iinva})] in TABLE~\ref{tab2}. Obviously, $I_{\textrm{inva}}[A_\alpha]$ is a certain value, where data set \emph{Exponent} has a really small value and \emph{Constant} a large value.
\begin{table}
\caption{\label{tab2}   $I_{\textrm{inva}}[A_\alpha]$ for different data sets}
\begin{tabular}{c| c| c| c}
\hline
\emph{Normal} & \emph{Exponent} &\emph{Uniform} & \emph{Constant}\\
\hline
0.087 & 0.031 & 0.198 & 0.907\\
\hline
\end{tabular}
\end{table}
Interestingly, for data set with normal distribution, which has zero average but arbitrary standard deviation, the calculated $I_{\textrm{inva}}[A_\alpha]$ is equal, just a fixed value about 0.087 in TABLE~\ref{tab2}. So, data set with any normal distribution, which can be easily transformed to zero average, has one value of "Scaling invariable Benford distance".

\begin{figure}
\includegraphics[width=0.5\textwidth]{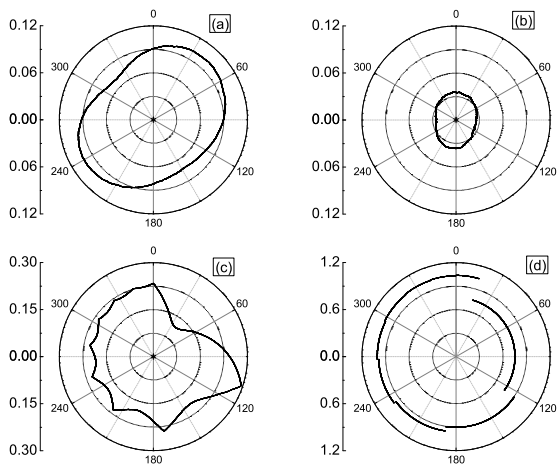}
\caption{\label{fig:fig2} As radial coordinate, $I[A_\alpha]$ [Eq.(~\ref{eq:Iaa})] varies with angular coordinate which is transformed from $\alpha\in[0,1)$ with uniform distribution. These four graphs are results of four data sets [TABLE.~\ref{tab1}], i.e.(a) \emph{Normal}, (b) \emph{Exponent}, (c) \emph{Uniform} and (d)  \emph{Constant}.} 
\end{figure}

\begin{figure}
\includegraphics[width=0.5\textwidth]{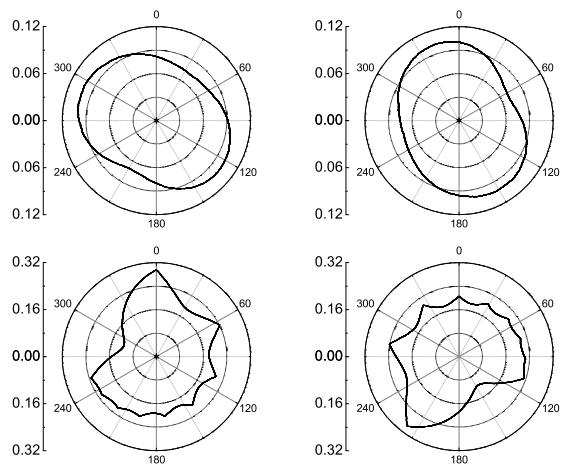}
\caption{\label{fig:fig3} As radial coordinate, $I[A_\alpha]$ [Eq.(~\ref{eq:Iaa})] varies with angular coordinate which is transformed from $\alpha\in[0,1)$ with uniform distribution. The upper two graphs show the results of data sets transformed from \emph{Normal} [TABLE.~\ref{tab1}], where all members multiply 2 (left graph) and 5 (right graph). The lower two graphs show results of data sets transformed similarly from \emph{Uniform} [TABLE.~\ref{tab1}].}
\end{figure}

Then, we calculate $I[A_\alpha]$ defined in Eq.(~\ref{eq:Iaa}) for any given $\alpha$, which is evenly distributed on interval $[0,1)$, and show the results for these four cases in FIG.~\ref{fig:fig2}, where the radial coordinate is $I[A_\alpha]$, and the angular coordinate is transformed from $\alpha$. Apparently, different data set has different figure feature, and such graph is called "Benford cyclic graph" here.
Moreover, if we change units or scales of one data set, we will get the same shape but rotation of the "Benford cyclic graph". For instance, in FIG.~\ref{fig:fig3}, the upper two graphs show results of data sets transformed from the above data set \emph{Normal}, where all members  multiply 2 (left graph) and 5 (right graph), correspondingly, polar angles of the graphs counterclockwise rotate through 108 and 252 degrees compared to graph (a) of FIG.~\ref{fig:fig2}. And the lower two graphs show results of data sets which are transformed similarly from \emph{Uniform}.

That is, both "Scaling invariable Benford distance" and "Benford cyclic graph" can be used to identify and classify data sets, and the former is easily to use while the latter gives much more information. 
\section{\label{sec:level4}The influence of the data size on first digit proportions}
\begin{figure}[b]
\begin{center}
\includegraphics[width=0.5\textwidth]{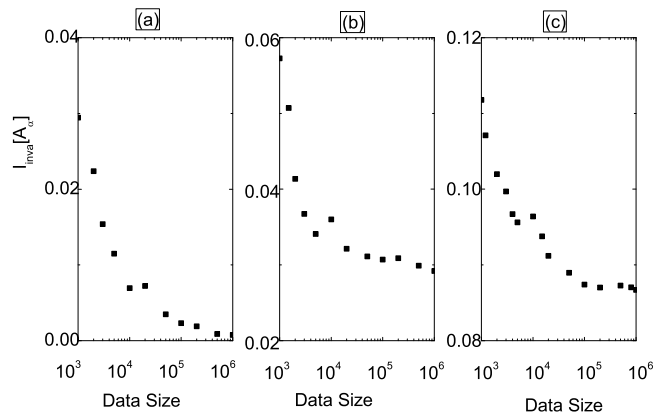}
\end{center}
\caption{\label{fig3} $I_{\textrm{inva}}[A_\alpha]$ [Eq.(~\ref{eq:Iinva})] varies with data size, which changes from $10^3$ to $10^6$. And the graphs are results of data sets, (a) $X$ defined in Eq.(~\ref{eqx}),(b) \emph{Exponent}, (c) \emph{Normal},both (b) and (c) from TABLE.~\ref{tab1}}
\end{figure}

Here, we explore the influence of the data size on first digit proportions by "Scaling invariable Benford distance". We calculate $I_{\textrm{inva}}[A_\alpha]$ for three data sets mentioned above, one is data set $X$ defined in Eq.(~\ref{eqx}), the other two are \emph{Exponent} and \emph{Normal} from TABLE.~\ref{tab1}. And results are showed 
in FIG.~\ref{fig3}, where data size changes from $10^3$ to $10^6$. Apparently, if the data size is less than $10^5$, $I_{\textrm{inva}}[A_\alpha]$ reduces greatly with the number increasing, and then it approximates to a value. Furthermore, for data set $X$, the approximation of $I_{\textrm{inva}}[A_\alpha]$ is zero
because its first digit proportions are consistent with \emph{Benford'law}, while the approximation is a nonzero value for other data sets. 
%
\section{\label{sec:level5} Conclusion}
According to the introduction of this paper, it's too risky to confirm that first digit proportions of a real-world data set are consistent with \emph{Benford'law}. For example, the data set of star distances of stars in Milky way~\cite{AEK} is supposed to fit the law extremely well. However, "Scaling invariable Benford distance" of this case is 0.0438, while it is just 0.0308 for data set \emph{Exponent} with the same data size (48111). Apparently, \emph{Exponent} can't be the case which satisfy the law because that the approximation of $I_{\textrm{inva}}[A_\alpha]$ is a nonzero number, so can't be the star distances. Such circumstances have been tackled through  "Scaling invariable Benford distance" firstly introduced in this paper.

Using this new quantity, we have analyzed some typical data sets, results of which show that  different data set has a much different value of "Scaling invariable Benford distance". We have also explored how the quantity varies with the data size, and found that $I_{\textrm{inva}}[A_\alpha]$ approximates a fixed value when the size is large enough, and the value is zero for data set $X$ [Eq.(~\ref{eqx})] which fits the digit law, nonzero number for other data sets. In addition, we have introduced "Benford cyclic graph" which can also identify and classify data sets as "Scaling invariable Benford distance", and in part~\ref{sec:level31} given a different proof that a large category of data sets satisfy \emph{Benford's law} in contrast to Hamming~\cite{HM}.

In general, "Scaling invariable Benford distance" and "Benford cyclic graph" can be used to analyze any data set, and thought as a statistical way, resulting in extending  research and applications of \emph{Benford's law}. For instance, one application is identifying the authenticity of the given data sets, which formerly must approximate the logarithmic law, but now can not. Another example is that we can estimate the distribution of a data set is as it is considered to be, such as normal distribution talked about in part~\ref{sec:level32}. Regarding these points, further studying need to be done for deeper meaning and broader application of "Scaling invariable Benford distance" and "Benford cyclic graph".

Moreover, our results raise an interesting question: we apply statistics to deal with real-world data sets which are usually with uncertainty, but statistics can give results just at the limiting case while the data size is always limited. So, this gap between limited data sets and the infinity in statistics, will bring with it what problems, we don't know. If there are such problems, the method like ours talked about in this paper, namely "Scaling invariable Benford distance" and "Benford cyclic graph", may be an inspiration because they are based upon the limited analysis.
\\

This work was supported by the National Magnetic Confinement Fusion Program of China (No.2014GB125004) and the National Natural Science Foundation of China (No.11575121).

\end{document}